\documentclass[twocolumn,english,aps,groupaddress,prb]{revtex4-2}

\usepackage[colorlinks=true,allcolors=blue]{hyperref}
\usepackage{array}
\usepackage[latin9]{inputenc}
\usepackage{amsmath}
\usepackage{amssymb}
\usepackage{graphicx}
\usepackage{babel}
\usepackage{mathrsfs}
\usepackage{amsfonts}
\usepackage{epstopdf}
\usepackage{multirow}
\usepackage{color}
\usepackage{natbib}
\usepackage{bm}

\usepackage{xcolor}

\begin{document}

\title{Euler--Chern Correspondence via Topological Superconductivity}
\author{Fan Yang}
\email{paullsc123@gmail.com}
\thanks{These two authors contributed equally}
\author{Xingyu Li}
\email{xyli22@mails.tsinghua.edu.cn}
\thanks{These two authors contributed equally}
\author{Chengshu Li}
\email{lichengshu272@gmail.com}
\affiliation{Institute for Advanced Study, Tsinghua University, Beijing, 100084, China}
\date{\today}

\begin{abstract}
    The Fermi sea topology is characterized by the Euler characteristics $\chi_F$. In this paper, we examine how $\chi_F$ of the metallic state is inhereted by the topological invariant of the superconducting state. We establish a correspondence between the Euler characteristic and the Chern number $C$ of $p$-wave topological superconductors without time-reversal symmetry in two dimensions. By rewriting the pairing potential $\Delta_{\bf k}=\Delta_1-i\Delta_2$ as a vector field ${\bf u}=(\Delta_1,\Delta_2)$, we found that $\chi_F=C$ when ${\bf u}$ and fermion velocity ${\bf v}$ can be smoothly deformed to be parallel or antiparallel on each Fermi surface.  We also discuss a similar correspondence between Euler characteristic and 3D winding number of time-reversal-invariant $p$-wave topological superconductors in three dimensions.
\end{abstract}

\maketitle

\section{Introduction} In the past few decades, it has been proven that topology plays an important role in physics~\cite{moessner_moore,Hasan10,Qi11,Bernevig,Sato17,Senthil15,Witten15,Wen17,Armitage18,Schnyder15,Tanaka12}. The topological invariant of a quantum system leads to quantized response functions and/or robust gapless states at the boundary. For example, the Chern number of an integer quantum Hall state determines the quantized Hall conductance and the number of chiral edge states \cite{Klitzing80,Thouless82}. Such topological invariants describe the twist of wavefunctions in the momentum space. Formally, the Hamiltonian defines a map from the momentum space to some target space, and the topological invariant is generally given by the homotopy group of this map. Most recent discussions on topological states have been concentrated on gapped systems \cite{Hasan10,Qi11,Bernevig,Sato17,Senthil15,Witten15,Wen17}, semimetals \cite{Armitage18}, and nodal superconductors \cite{Schnyder15}.

Different from the topology related to wavefunctions, there exists another type of geometric topology in Fermi liquids. The Fermi sea as a manifold can have complicated structures, and its topology is characterized by the Euler characteristic $\chi_F$  \footnote{Mathematically, the Euler characteristic of a $d$-dimensional Fermi sea is defined as the alternating sum of its Betti numbers $b_k$, $\chi_F=\sum_{k=0}^d(-1)^k b_k$. See also \cite{TopoPhys}}. The Euler characteristic changes when the Fermi level passes through a critical point of the energy dispersion.
The change of $\chi_F$ is accompanied by the famous Lifshitz transitions \cite{Lifshitz60,Volovik17}. 
In principle, one can always map out the entire Fermi surface to obtain $\chi_F$, but only recently was it realized that $\chi_F$ can be measured directly.
An important breakthrough is the prediction that the Euler characteristic $\chi_F$ determines a quantized nonlinear conductance in $d$-dimensional ballistic metals \cite{Kane22}. This phenomenon can be viewed as a generalization of the quantized Landauer conductance in one dimension (1D) \cite{Landauer57}. It also reveals a deep connection between Fermi sea topology and quantized transport properties.
This sparked a series of efforts in directly detecting $\chi_F$, including probing multipartite entanglement entropy \cite{Tam22}, measuring quantized response in ultracold Fermi gases \cite{Yang22QNL,PFZhang23}, and utilizing Andreev state transport \cite{Tam23,tam2023topological}.

\section{Summary of Results} In this paper, we discover a relation between the Euler characteristic and the topological invariants of $p$-wave  topological superconductors in two (2D) and three dimensions (3D). We provide a condition under which the Euler characteristic of the metallic state and the topological invariant of the superconductor are equal. When this condition is satisfied, the Majorana states at the boundary of the superconductor can be used to measure the Euler characteristic.

\begin{figure}
	\centering
	\includegraphics[width=\columnwidth]{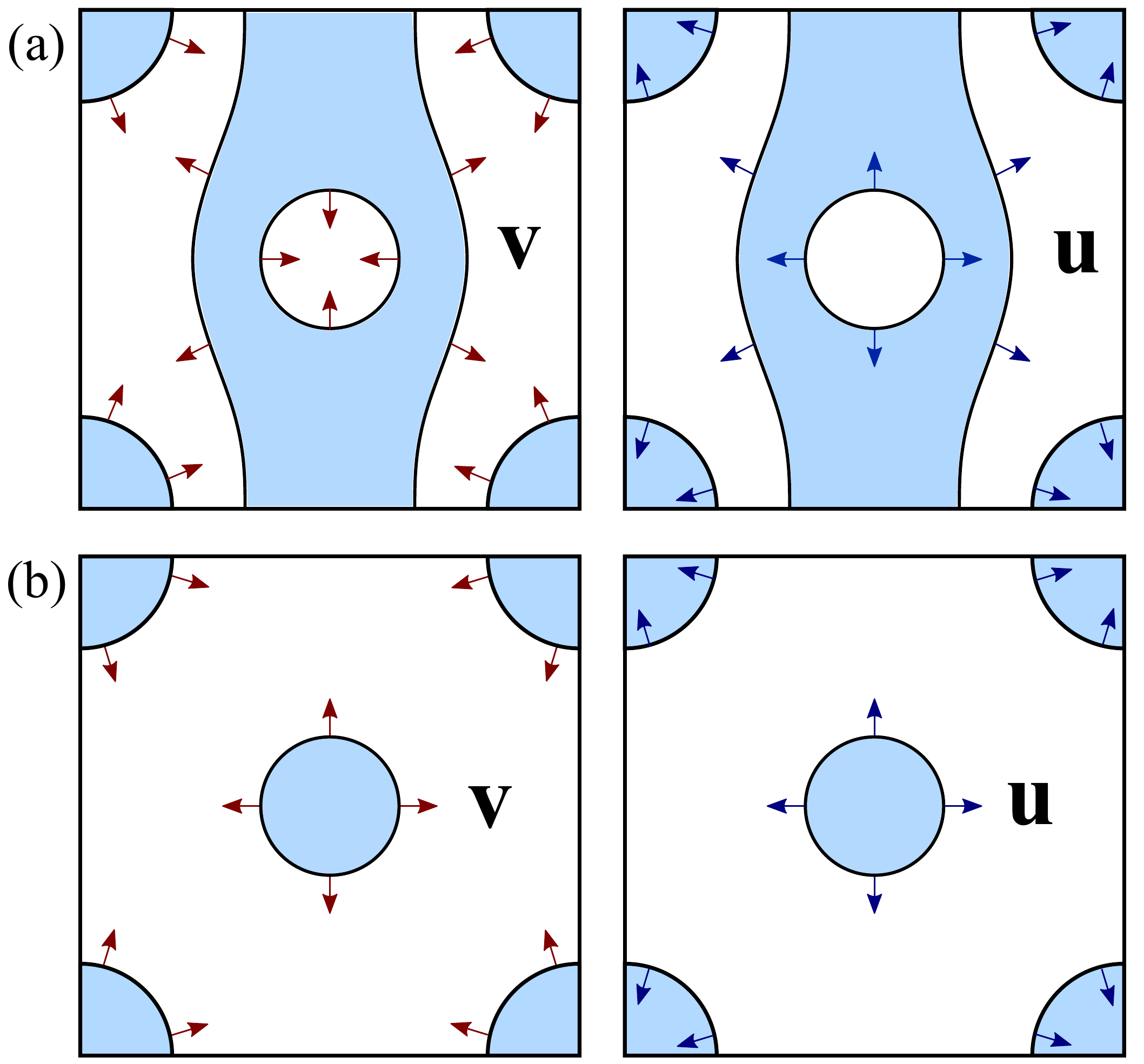}
	\caption{Illustration of spinless Fermi seas with (a) $\chi_F=0$ and (b) $\chi_F=2$, respectively. The red arrows on the left panel represent the fermion velocity ${\bf v}$ on the Fermi surface. The blue arrows on the right panel represent the pairing vector field ${\bf u}$ of a $p-ip$ superconductor. On each Fermi surface, ${\bf v}$ and ${\bf u}$ can be smoothly deformed without vanishing to be parallel or antiparallel, giving $\chi_F=C$.}
	\label{fig:FS}
\end{figure}

In 2D the pairing potential must break time-reversal symmetry so that the resultant topological superconductor is characterized by an integer that is the Chern number $C$. For simplicity, we first focus on the spinless case.
When the following condition is satisfied,  $\chi_F$ and $C$ are equal.

{\bf Condition:} We write the pairing potential $\Delta_{\bf k}=\Delta_1-i\Delta_2$ as a vector field ${\bf u}=(\Delta_1,\Delta_2)$. In the weak pairing limit $|\Delta|\to0$, if ${\bf u}$ can be smoothly deformed without vanishing to be parallel or antiparallel to the fermion velocity ${\bf v}=\nabla_{\bf k}\epsilon_{\bf k}/\hbar$ on each Fermi surface, we have 
\begin{equation}\label{eq:equal}
    \chi_F=C,
\end{equation} 
where $\epsilon_{\bf k}$ is the energy dispersion of the metallic state. 

To satisfy the above condition, one would need a $p$-wave pairing potential that has the proper chirality on each Fermi surface. 
In Fig. \ref{fig:FS}, we present two examples where this condition can be met by a simple $p-ip$ pairing. In this case, the number of chiral Majorana edge modes in the topological superconductor is given by $\chi_F$. By measuring the quantized thermal Hall conductance given by the Majorana edge modes \cite{Kitaev06QSL,Kane97,Cappelli02}, one can probe $\chi_F$ through $C$.
A simple $p\pm ip$ pairing cannot always satisfy the condition. However, if an additional inversion symmetry is present, $p\pm ip$ pairing always leads to $\chi_F\equiv C\pmod2$. We note a related result in Refs. \cite{Sato09,Sato10}, where it was shown that the Chern number and a different topological invariant, which is the number of Fermi surfaces, have the same parity for inversion symmetric odd-parity superconductors.

A similar relation exists in 3D if the system respects time-reversal symmetry. In this case, we introduce a time-reversal-invariant $p$-wave pairing $\Delta_{\bf k}={\bf u}\cdot{\bm\sigma}i\sigma_y$, where ${\bm \sigma}=(\sigma_x,\sigma_y,\sigma_z)$ are Pauli matrices in the spin space and ${\bf u}$ is odd in ${\bf k}$. The resultant topological superconductor is characterized by the 3D winding number $N_w\in \mathbb{Z}$. Due to the presence of both spin components, $\chi_F$ is always an even integer. 
When ${\bf u}$ and fermion velocity ${\bf v}$ can be smoothly deformed to become parallel with each other on each Fermi surface, we have 
\begin{equation}
   \frac{\chi_F}{2}=N_w. 
\end{equation}

Similar to 2D, a simple time-reversal-invariant $p$-wave pairing cannot always satisfy the condition above. But with inversion symmetry, it always leads to $\chi_F/2\equiv N_w\pmod2$, which was given previously in Ref. \cite{Sato09,Sato10}.  

The above correspondences are robust against weak interactions, since both Fermi surfaces and topological superconductors remain well-defined.
In the following, we mathematically establish the conditions given above.

\section{Euler--Chern correspondence in 2D}
The connection between the Euler characteristic and the Chern number may seem surprising at first, since the Euler characteristic is defined for gapless Fermi liquids, while the Chern number is defined for a fully gapped system without a Fermi surface. In 2D, the Euler characteristic is given by the number of electron-like Fermi surfaces minus the number of hole-like Fermi surfaces, while open Fermi surfaces do not contribute to $\chi_F$. On the other hand, the Chern number is given by the integration of Berry curvature over the entire Brillouin zone. However, in the weak pairing limit, the Berry curvature in a superconductor is concentrated near the Fermi surface. And the sign of the Berry curvature near the Fermi surface depends on if the Fermi surface is electron-like, hole-like, or open. In this way, the information of Fermi sea topology is encoded into the Chern number.

To generally establish the relation between $\chi_F$ and $C$, we express them as the winding number on the Fermi surface of two different vector fields ${\bf v}$ and ${\bf u}$, respectively. 
${\bf v}$ is given by the fermion velocity, and ${\bf u}$ is given by the pairing potential $\Delta_{\bf k}$.
Let us first consider the spinless case.
At each point on the Fermi surface, the fermion velocity ${\bf v}=\nabla_{\bf k}\epsilon_{\bf k}/\hbar$ is always perpendicular to the Fermi surface and pointing away from the Fermi sea. With the help of Poincare--Hopf theorem \cite{TopoDiff}, one can write the Euler characteristic $\chi_F$ as the sum of the winding numbers of ${\bf v}$ on all Fermi surfaces
\begin{equation}\label{eq:winding}
    \chi_F=\sum_\alpha w_\alpha({\bf v}),
\end{equation}
where $w_\alpha({\bf v})$ is the winding number of ${\bf v}$ on Fermi surface $S_\alpha$. Each electron-like, hole-like, and open Fermi surface has winding number $+1$, $-1$, and $0$, thus contributing $+1$, $-1$, and $0$ to $\chi_F$.

Let us consider a lattice version of $p-ip$ superconductor with pairing potential $\Delta_{\bf k}=\Delta_0(\sin{\bf k\cdot a_1}-i\sin{\bf k\cdot a_2})$, where ${\bf a_{1,2}}$ are lattice basis vectors \cite{Read00,Kallin16}. The Bogoliubov--de Gennes (BdG) Hamiltonian can be written as
\begin{equation}
    H({\bf k})=(\epsilon_{\bf k}-\mu)\tau_z+\Delta_0\sin {\bf k\cdot a_1}\tau_x+\Delta_0\sin {\bf k\cdot a_2}\tau_y,
\end{equation}
where $\tau_{x,y,z}$ are Pauli matrices in the Nambu space and $\mu$ is the chemical potential. The pairing breaks time-reversal symmetry and the resultant topological superconductor belongs to the D class, which is characterized by the Chern number \cite{Schnyder08,Kitaev09,Chiu16}.
$H({\bf k})$ defines a map from the Brillouin zone to a unit sphere given by $\hat{\bf h}={\bf h}/|{\bf h}|$ with ${\bf h}=(\Delta_0\sin {\bf k\cdot a_1}, \Delta_0\sin {\bf k\cdot a_2}, \mu-\epsilon_{\bf k})$. 
The Chern number $C$ equals to the degree of this map
\begin{equation}\label{eq:Chern}
    C=\frac{1}{4\pi}\int_\text{1BZ}d^2k\,\hat{\bf h}\cdot\left(\frac{\partial\hat{\bf h}}{\partial k_x}\times\frac{\partial\hat{\bf h}}{\partial k_y}\right),
\end{equation}
which is the number of times that $\hat{\bf h}$ covers the unit sphere. 
In the weak pairing limit $\Delta_0/\mu\ll1$, the Berry curvature given by the integrand of Eq. (\ref{eq:Chern}) is concentrated near the Fermi surface. Except for a thin shell near each Fermi surface, $\hat{\bf h}$ points along $\hat{z}$ inside and $-\hat{z}$ outside the Fermi sea, respectively. On each Fermi surface, $h_z=\mu-\epsilon_{\bf k}=0$ and ${\bf h}$ is reduced to a two-dimensional vector field ${\bf u}=(\Delta_0\sin{\bf k\cdot a_1},\Delta_0\sin {\bf k\cdot a_2})$.
Therefore, in the weak pairing limit, Eq. (\ref{eq:Chern}) becomes \cite{TopoPhys}
\begin{equation}\label{eq:ChernWinding}
    C=\sum_\alpha w_\alpha({\bf u}),
\end{equation}
where 
\begin{equation}
    w_\alpha({\bf u})=\frac{1}{2\pi}\oint_{{S}_\alpha}(\hat{u}_xd\hat{u}_y-\hat{u}_yd\hat{u}_x)
\end{equation}
is the winding number of $\hat{\bf u}\equiv{\bf u}/|{\bf u}|$ on the $\alpha$th Fermi surface $S_\alpha$ with its orientation induced by the Fermi sea.  

Eqs. (\ref{eq:winding}) and (\ref{eq:ChernWinding}) suggest that if $w_\alpha({\bf v})=w_\alpha({\bf u})$ on each Fermi surface $S_\alpha$, we will have $\chi_F=C$. This is equivalent to the condition that the pairing vector field ${\bf u}$ can be smoothly deformed without vanishing to be parallel or antiparallel to fermion velocity ${\bf v}$ on each Fermi surface. This is the condition given in the {\it Results}.

If this condition cannot be satisfied with a simple $p\pm ip$ pairing, we would require that Cooper pairs be formed by electrons from the same Fermi surface in the $p$-wave channel with the proper chirality, which requires special engineering of the pairing potential. In the weak pairing limit, one can write a BdG-like Hamiltonian near each Fermi surface. By choosing the proper chirality of $p$-wave pairing potential near each Fermi surface, we can have $w_\alpha({\bf u})=w_\alpha({\bf v})$. 

Physically, if simple $p\pm ip$ pairings cannot satisfy the condition, we can still have $\chi_F\equiv C\pmod2$ if inversion symmetry is present.
To see this, we extend both vector fields to the entire Brillouin zone and analyze their zeros inside the Fermi sea. 

\begin{figure}
    \centering
    \includegraphics[width=\columnwidth]{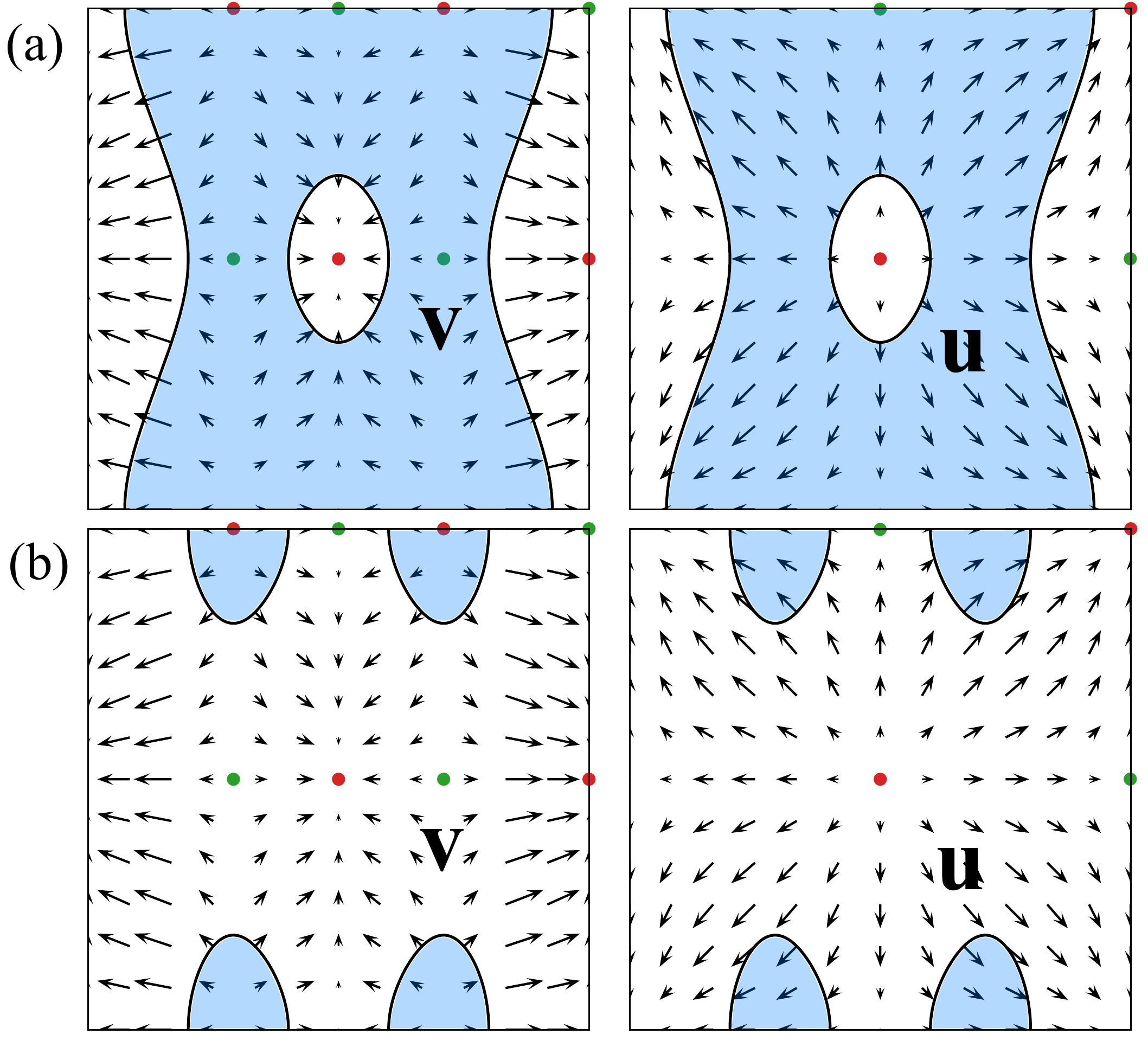}
    \caption{Illustration of spinless Fermi sea with dispersion $\epsilon_{\bf k}=-\cos(k_x)+\cos(k_y)+\cos(2k_x)$. (a) $\mu=0.5$, $\chi_F=-1$, and $C=-1$. (b) $\mu=-1.5$, $\chi_F=2$, and $C=0$. The left column shows the fermion velocity and the right column pairing field ${\bf u}$ extended to the entire Brillouin zone. Red and green dots label the zeros with indices $+1$ and $-1$, respectively.}
    \label{fig:field}
\end{figure}

The zeros are defined as the points where the vector field vanishes \footnote{In the following, we assume all zeros are isolated. When there are lines of zeros, one can always smoothly change the vector field by an infinitesimal amount such that all zeros are isolated.}. Each zero can be assigned an index $\nu_i$, which is the winding number of the vector field on a small counterclockwise oriented circle enclosing the $i$th zero. Zeros with indices $+1$ and $-1$ can be viewed as vortices and antivortices of the vector field, respectively.
Since the winding number of a vector field at the boundary of a manifold equals to the sum of the indices in its interior, we have
\begin{equation}\label{eq:index}
    \chi_F=\sum_\alpha w_\alpha({\bf v})=\sum_{i\in\text{FS}}\nu_i({\bf v}),
\end{equation}
\begin{equation}
    C=\sum_\alpha w_\alpha({\bf u})=\sum_{j\in\text{FS}}\nu_j({\bf u}),
\end{equation}
where $\nu_i({\bf v})$ and $\nu_j({\bf u})$ are the indices of ${\bf v}$ and ${\bf u}$ fields, respectively. We use different subscripts $i$ and $j$ to remind the reader that the number of zeros for ${\bf v}$ and ${\bf u}$ fields could be different.
On the right hand side, the summation is over all zeros inside the Fermi sea (FS)~\footnote{Eq. (\ref{eq:index}) is equivalent to the Morse theory formulation of $\chi_F$. $\nu_i({\bf v})$ equals to the sign of the Hessian of $\epsilon_{\bf k}$ at each critical point.} (Fig.~\ref{fig:field}). 

We compare the indices of zeros of ${\bf v}$ and ${\bf u}$ inside the Fermi sea.
For inversion symmetric systems, both ${\bf v}$ and ${\bf u}$ are odd under inversion, i.e., ${\bf v}({\bf k})=-{\bf v}({\bf -k+G})$ and ${\bf u}({\bf k})=-{\bf u}({\bf -k+G})$. All time-reversal invariant momenta (TRIM) $\Gamma_{n_1n_2}=\frac{n_1}{2}{\bf b_2}+\frac{n_2}{2}{\bf b_2}$ are inversion centers and therefore also the zeros, where ${\bf b_1}$ and ${\bf b_2}$ are reciprocal lattice basis vectors. Along any direction across the inversion center, the vector field must change to its opposite direction. Thus, the index at each TRIM for both vector fields must be odd \cite{Hatcher}. For vector fields odd under inversion, other zeros must appear in pairs at ${\bf k}$ and ${\bf -k+G}$. These zeros must have the same index and both appear inside or outside the Fermi sea. Therefore, we have
\begin{equation}
    \sum_{i\in\text{FS}}\nu_i({\bf v})\equiv\sum_{j\in\text{FS}}\nu_j({\bf u}) \pmod 2, 
\end{equation}
i.e.,
\begin{equation}\label{eq:correspondence}
    \chi_F\equiv C \pmod 2.
\end{equation}
Note that Eq. (\ref{eq:correspondence}) holds for any odd-parity pairing potential, not just the chiral $p$-wave pairing chosen above.

When both spin components are considered, without spin--orbit coupling, we demand a time-reversal-breaking $p$-wave pairing with spin $U(1)$ rotational symmetry. 
For example, we can have each spin component paired in the same chiral $p$-wave channel, given by the pairing potential $\Delta_{\bf k}=\Delta_0(\sin{\bf k\cdot a_1}-i\sin{\bf k\cdot a_2})\sigma_0$, where $\sigma_0$ is the identity matrix in the spin space. The pairing potential explicitly breaks time-reversal symmetry. The resultant superconducting state is the 2D analog of the He-3 A phase, with the spin rotational axis along the $y$-direction \cite{Volovik88,Leggett75,Volovik,Vollhardt}. More generally, one can choose an arbitrary spin rotational axis ${\bf \hat{n}}$ for the spinful case, and the corresponding pairing potential becomes
\begin{equation}\label{eq:pairing}
\begin{split}
    \Delta_{\bf k}&=\Delta_0(\sin{\bf k\cdot a_1}-i\sin{\bf k\cdot a_2}){\bf \hat{n}}\cdot{\bm \sigma}i\sigma_y\\
    &\equiv(\Delta_1-i\Delta_2){\bf \hat{n}}\cdot{\bm \sigma}i\sigma_y,
\end{split}
\end{equation}
with ${\bf u}$ again defined as ${\bf u}=(\Delta_1,\Delta_2)$.
One can also introduce a weak spin--orbit coupling that does not change $\chi_F$. In this case, the topological superconductor is still classified by the Chern number and its value does not change as long as the superconducting gap remains open. Therefore, our results still apply.
When multiple bands are present at the Fermi surface, we would only consider intraband pairing that break time-reversal symmetry. That is the pairing potential is of the form of Eq. (\ref{eq:pairing}) for each band. If for each band, ${\bf u}$ satisfies the condition, we have $\chi_F=C$. Otherwise, we have $\chi_F\equiv C \pmod 2$ when inversion symmetry is present.

\section{Generalization to 3D}
Next, we generalize our result to 3D, where each Fermi surface with genus $g$ contributes $1-g$ to the Euler characteristic $\chi_F$ of the Fermi sea. 
We require the metallic state respect time-reversal symmetry of spin-$1/2$ fermions.  
By introducing a time-reversal invariant $p$-wave pairing, we can convert the metallic state to a topological superconductor of the DIII class, which is classified by an integer. 

The pairing potential has the form $\Delta_{\bf k}={\bf u}\cdot{\bm\sigma}i\sigma_y$, where ${\bm\sigma}=(\sigma_x,\sigma_y,\sigma_z)$ are Pauli matrices in the spin space, and ${\bf u}=(\Delta_0\sin{\bf k\cdot a_1},\Delta_0\sin{\bf k\cdot a_2},\Delta_0\sin{\bf k\cdot a_3})$ is the pairing vector field in 3D. In the continuous limit, this corresponds to the He-3 B phase \cite{Leggett75,Volovik,Vollhardt}. Without spin--orbit coupling, the Hamiltonian can be written as
\begin{equation}
\begin{split}
    H({\bf k})={}&(\epsilon_{\bf k}-\mu)\tau_z-\Delta_0\sin{\bf k\cdot a_1}\tau_x\sigma_z\\
    &-\Delta_0\sin{\bf k\cdot a_2}\tau_y+\Delta_0\sin{\bf k\cdot a_3}\tau_x\sigma_x.
\end{split}
\end{equation}
$H({\bf k})$ defines a map from the 3D Brillouin zone to a unit 3-sphere given by $\hat{\bf h}={\bf h}/|{\bf h}|$, with ${\bf h}=({\bf u},\mu-\epsilon_k)$. This map is classified by the homotopy group $\pi_3(S^3)$, and its topological invariant is the 3D winding number $N_w$. When spin--orbit coupling is taken into account, the topological invariant is still the 3D winding number, although the homotopy group becomes $\pi_3(U(2))$.

In complete analogy to the 2D case, we can rewrite $N_w$ as the sum of 2D winding numbers of ${\bf u}$ on each Fermi surface in the weak pairing limit.
Similarly, the Euler characteristic is given by the sum of the winding numbers of fermion velocity ${\bf v}$ on each Fermi surface. Due to the presence of both spins, $\chi_F$ is an even integer.
When ${\bf u}$ can be smoothly deformed without vanishing to be parallel to ${\bf v}$, we have $\chi_F/2=N_w$.
If this condition cannot be met, in the presence of inversion symmetry, we have $\chi_F/2\equiv N_w \pmod2$. When multiple bands are present, we would only consider intraband pairing, with the pairing potential given by $\Delta_k$ for each band.

\section{Experimental implications}
We discuss two experimental implications of the above correspondences.

\begin{figure}
    \centering
    \includegraphics[width=\columnwidth]{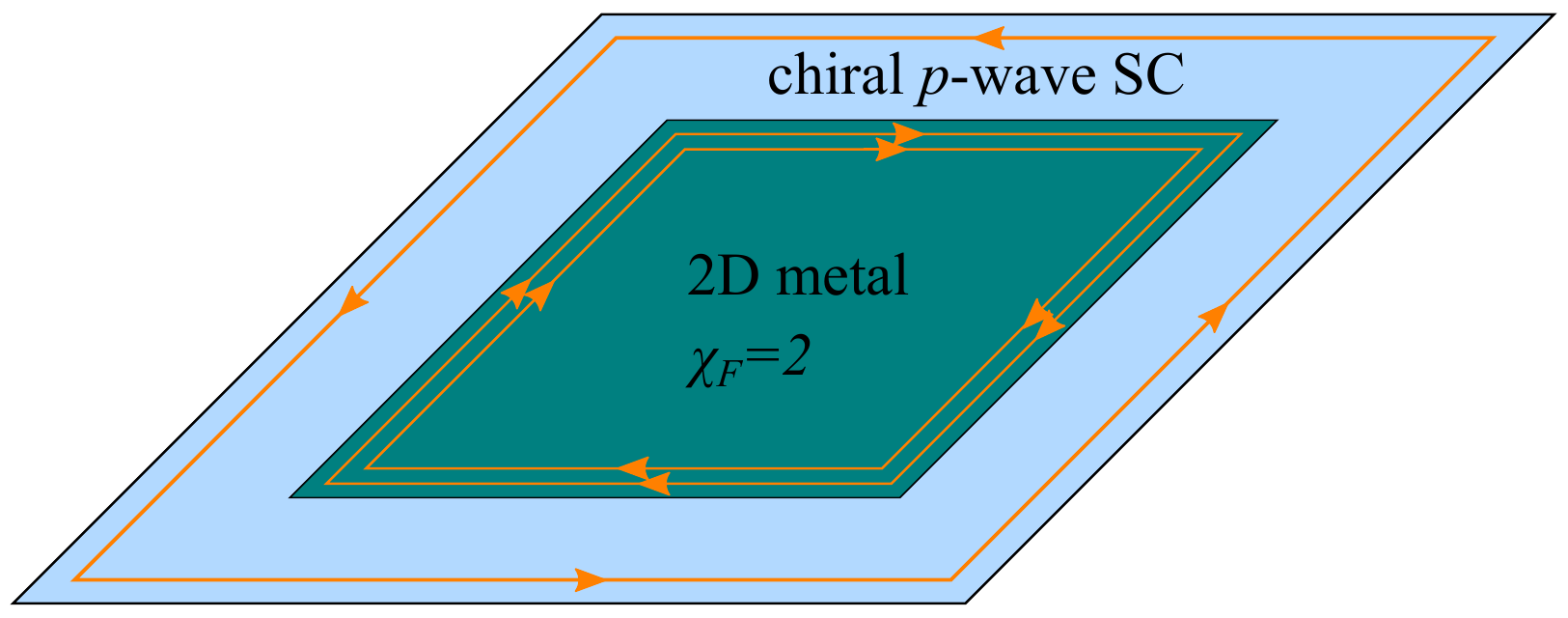}
    \caption{We induce chiral $p$-wave superconductivity to a 2D metallic sample by proximity effect. The number of chiral Majorana mode on the sample depends on its Euler characteristic. The number and chirality of the Majorana modes in the sample can be different from that of the substrate.}
    \label{fig:proximity}
\end{figure}

To verify the correspondence between Euler characteristic and Chern number, we utilize the superconducting proximity effect \cite{Fu08,Abrikosov}. Let us consider spinless fermions. In general, Cooper pairs would not always form in a time-reversal-breaking $p$-wave channel. However, we can induce such pairings by depositing the 2D metallic sample onto a chiral $p$-wave superconducting substrate. If the  condition is satisfied, then the number and chirality of the Majorana edge modes in the sample is given by $\chi_F$. If the condition is not satisfied, with inversion symmetry the number of chiral Majorana edge modes is given by $\chi_F$ modulo two. In either case, the number and chirality of Majorana modes in the sample can be different from that in the substrate (Fig. \ref{fig:proximity}).

The relation between Euler characteristic and topological invariants of superconductors also suggests that Lifshitz transitions in the metallic phase can lead to topological phase transitions in the superconducting phase in both 2D and 3D \cite{Read00,Yang19,Yang21,Zhou22,Yerin22}. For example, by applying pressure on a $p$-wave topological superconductor, one should be able to observe topological phase transitions marked by the change of Majorana edge modes. This topological phase transition is due to the change of Fermi sea topology of the metallic state by the applied pressure \cite{Lifshitz60}. When ${\bf u}$ and ${\bf v}$ satisfy the conditions given above, the superconductor undergoes a topological phase transition whenever $\chi_F$ changes its value. If the condition is not satisfied, with inversion symmetry topological phase transitions happen when $\chi_F$ changes its parity.

\section{Conclusion and Discussions}
In this paper, we have established the correspondence between the topological invariant of superconductors and the Euler characteristic of the normal state Fermi sea. The key to establishing this correspondence is to express the Euler characteristic and the topological invariant of superconductors as the winding numbers of the fermion velocity ${\bf v}$ and the pairing field ${\bf u}$ on the Fermi surface, respectively.
The way to express topological invariants as properties on the Fermi surface is reminiscent of the introduction of Fermi surface topological invariants for time-reversal invariant superconductors \cite{Qi10} and the characterization of Floquet topological phases with band-inversion surface properties \cite{Zhang20}.

Our work reveals a connection between two seemingly unrelated topological invariants in two physical systems with drastically different properties.
One related question that is interesting to study in the future is whether there exists similar correspondence between other topological invariants and what is the physical mechanism to connect them. Answering this question will help us understand the relations between different topological phases and may eventually lead to a more unified understanding of topology in physics. 

\begin{acknowledgments}
We thank Hui Zhai, Yingfei Gu, Pengfei Zhang, and Zhong Wang for helpful discussions. This project is supported by China Postdoctoral Science Foundation (Grant No. 2022M711868).
F.Y. and C.L. are  supported by Chinese International Postdoctoral Exchange Fellowship Program (Talent-introduction Program) and Shuimu Tsinghua Scholar Program at Tsinghua University.
\end{acknowledgments}

\bibliography{biblio}

\begin{thebibliography}{50}%
\makeatletter
\providecommand \@ifxundefined [1]{%
 \@ifx{#1\undefined}
}%
\providecommand \@ifnum [1]{%
 \ifnum #1\expandafter \@firstoftwo
 \else \expandafter \@secondoftwo
 \fi
}%
\providecommand \@ifx [1]{%
 \ifx #1\expandafter \@firstoftwo
 \else \expandafter \@secondoftwo
 \fi
}%
\providecommand \natexlab [1]{#1}%
\providecommand \enquote  [1]{``#1''}%
\providecommand \bibnamefont  [1]{#1}%
\providecommand \bibfnamefont [1]{#1}%
\providecommand \citenamefont [1]{#1}%
\providecommand \href@noop [0]{\@secondoftwo}%
\providecommand \href [0]{\begingroup \@sanitize@url \@href}%
\providecommand \@href[1]{\@@startlink{#1}\@@href}%
\providecommand \@@href[1]{\endgroup#1\@@endlink}%
\providecommand \@sanitize@url [0]{\catcode `\\12\catcode `\$12\catcode
  `\&12\catcode `\#12\catcode `\^12\catcode `\_12\catcode `\%12\relax}%
\providecommand \@@startlink[1]{}%
\providecommand \@@endlink[0]{}%
\providecommand \url  [0]{\begingroup\@sanitize@url \@url }%
\providecommand \@url [1]{\endgroup\@href {#1}{\urlprefix }}%
\providecommand \urlprefix  [0]{URL }%
\providecommand \Eprint [0]{\href }%
\providecommand \doibase [0]{https://doi.org/}%
\providecommand \selectlanguage [0]{\@gobble}%
\providecommand \bibinfo  [0]{\@secondoftwo}%
\providecommand \bibfield  [0]{\@secondoftwo}%
\providecommand \translation [1]{[#1]}%
\providecommand \BibitemOpen [0]{}%
\providecommand \bibitemStop [0]{}%
\providecommand \bibitemNoStop [0]{.\EOS\space}%
\providecommand \EOS [0]{\spacefactor3000\relax}%
\providecommand \BibitemShut  [1]{\csname bibitem#1\endcsname}%
\let\auto@bib@innerbib\@empty
\bibitem [{\citenamefont {Moessner}\ and\ \citenamefont
  {Moore}(2021)}]{moessner_moore}%
  \BibitemOpen
  \bibfield  {author} {\bibinfo {author} {\bibfnamefont {R.}~\bibnamefont
  {Moessner}}\ and\ \bibinfo {author} {\bibfnamefont {J.~E.}\ \bibnamefont
  {Moore}},\ }\href {https://doi.org/10.1017/9781316226308} {\emph {\bibinfo
  {title} {Topological Phases of Matter}}}\ (\bibinfo  {publisher} {Cambridge
  University Press},\ \bibinfo {year} {2021})\BibitemShut {NoStop}%
\bibitem [{\citenamefont {Hasan}\ and\ \citenamefont {Kane}(2010)}]{Hasan10}%
  \BibitemOpen
  \bibfield  {author} {\bibinfo {author} {\bibfnamefont {M.~Z.}\ \bibnamefont
  {Hasan}}\ and\ \bibinfo {author} {\bibfnamefont {C.~L.}\ \bibnamefont
  {Kane}},\ }\bibfield  {title} {\bibinfo {title} {Colloquium: Topological
  insulators},\ }\href {https://doi.org/10.1103/RevModPhys.82.3045} {\bibfield
  {journal} {\bibinfo  {journal} {Rev. Mod. Phys.}\ }\textbf {\bibinfo {volume}
  {82}},\ \bibinfo {pages} {3045} (\bibinfo {year} {2010})}\BibitemShut
  {NoStop}%
\bibitem [{\citenamefont {Qi}\ and\ \citenamefont {Zhang}(2011)}]{Qi11}%
  \BibitemOpen
  \bibfield  {author} {\bibinfo {author} {\bibfnamefont {X.-L.}\ \bibnamefont
  {Qi}}\ and\ \bibinfo {author} {\bibfnamefont {S.-C.}\ \bibnamefont {Zhang}},\
  }\bibfield  {title} {\bibinfo {title} {Topological insulators and
  superconductors},\ }\href {https://doi.org/10.1103/RevModPhys.83.1057}
  {\bibfield  {journal} {\bibinfo  {journal} {Rev. Mod. Phys.}\ }\textbf
  {\bibinfo {volume} {83}},\ \bibinfo {pages} {1057} (\bibinfo {year}
  {2011})}\BibitemShut {NoStop}%
\bibitem [{\citenamefont {Bernevig}\ and\ \citenamefont
  {Hughes}(2013)}]{Bernevig}%
  \BibitemOpen
  \bibfield  {author} {\bibinfo {author} {\bibfnamefont {B.~A.}\ \bibnamefont
  {Bernevig}}\ and\ \bibinfo {author} {\bibfnamefont {T.~L.}\ \bibnamefont
  {Hughes}},\ }\href
  {https://press.princeton.edu/books/ebook/9781400846733/topological-insulators-and-topological-superconductors}
  {\emph {\bibinfo {title} {{Topological insulators and topological
  superconductors}}}}\ (\bibinfo  {publisher} {Princeton University Press},\
  \bibinfo {year} {2013})\BibitemShut {NoStop}%
\bibitem [{\citenamefont {Sato}\ and\ \citenamefont {Ando}(2017)}]{Sato17}%
  \BibitemOpen
  \bibfield  {author} {\bibinfo {author} {\bibfnamefont {M.}~\bibnamefont
  {Sato}}\ and\ \bibinfo {author} {\bibfnamefont {Y.}~\bibnamefont {Ando}},\
  }\bibfield  {title} {\bibinfo {title} {Topological superconductors: a
  review},\ }\href {https://doi.org/10.1088/1361-6633/aa6ac7} {\bibfield
  {journal} {\bibinfo  {journal} {Rep. Prog. Phys.}\ }\textbf {\bibinfo
  {volume} {80}},\ \bibinfo {pages} {076501} (\bibinfo {year}
  {2017})}\BibitemShut {NoStop}%
\bibitem [{\citenamefont {Senthil}(2015)}]{Senthil15}%
  \BibitemOpen
  \bibfield  {author} {\bibinfo {author} {\bibfnamefont {T.}~\bibnamefont
  {Senthil}},\ }\bibfield  {title} {\bibinfo {title} {Symmetry-protected
  topological phases of quantum matter},\ }\href
  {https://doi.org/10.1146/annurev-conmatphys-031214-014740} {\bibfield
  {journal} {\bibinfo  {journal} {Annu. Rev. Condens. Matter Phys.}\ }\textbf
  {\bibinfo {volume} {6}},\ \bibinfo {pages} {299} (\bibinfo {year}
  {2015})}\BibitemShut {NoStop}%
\bibitem [{\citenamefont {Witten}(2016)}]{Witten15}%
  \BibitemOpen
  \bibfield  {author} {\bibinfo {author} {\bibfnamefont {E.}~\bibnamefont
  {Witten}},\ }\bibfield  {title} {\bibinfo {title} {Three lectures on
  topological phases of matter},\ }\href
  {https://doi.org/10.1393/ncr/i2016-10125-3} {\bibfield  {journal} {\bibinfo
  {journal} {Riv. Nuovo Cim.}\ }\textbf {\bibinfo {volume} {39}},\ \bibinfo
  {pages} {313} (\bibinfo {year} {2016})}\BibitemShut {NoStop}%
\bibitem [{\citenamefont {Wen}(2017)}]{Wen17}%
  \BibitemOpen
  \bibfield  {author} {\bibinfo {author} {\bibfnamefont {X.-G.}\ \bibnamefont
  {Wen}},\ }\bibfield  {title} {\bibinfo {title} {Colloquium: Zoo of
  quantum-topological phases of matter},\ }\href
  {https://doi.org/10.1103/RevModPhys.89.041004} {\bibfield  {journal}
  {\bibinfo  {journal} {Rev. Mod. Phys.}\ }\textbf {\bibinfo {volume} {89}},\
  \bibinfo {pages} {041004} (\bibinfo {year} {2017})}\BibitemShut {NoStop}%
\bibitem [{\citenamefont {Armitage}\ \emph {et~al.}(2018)\citenamefont
  {Armitage}, \citenamefont {Mele},\ and\ \citenamefont
  {Vishwanath}}]{Armitage18}%
  \BibitemOpen
  \bibfield  {author} {\bibinfo {author} {\bibfnamefont {N.~P.}\ \bibnamefont
  {Armitage}}, \bibinfo {author} {\bibfnamefont {E.~J.}\ \bibnamefont {Mele}},\
  and\ \bibinfo {author} {\bibfnamefont {A.}~\bibnamefont {Vishwanath}},\
  }\bibfield  {title} {\bibinfo {title} {Weyl and {D}irac semimetals in
  three-dimensional solids},\ }\href
  {https://doi.org/10.1103/RevModPhys.90.015001} {\bibfield  {journal}
  {\bibinfo  {journal} {Rev. Mod. Phys.}\ }\textbf {\bibinfo {volume} {90}},\
  \bibinfo {pages} {015001} (\bibinfo {year} {2018})}\BibitemShut {NoStop}%
\bibitem [{\citenamefont {Schnyder}\ and\ \citenamefont
  {Brydon}(2015)}]{Schnyder15}%
  \BibitemOpen
  \bibfield  {author} {\bibinfo {author} {\bibfnamefont {A.~P.}\ \bibnamefont
  {Schnyder}}\ and\ \bibinfo {author} {\bibfnamefont {P.~M.~R.}\ \bibnamefont
  {Brydon}},\ }\bibfield  {title} {\bibinfo {title} {Topological surface states
  in nodal superconductors},\ }\href
  {https://doi.org/10.1088/0953-8984/27/24/243201} {\bibfield  {journal}
  {\bibinfo  {journal} {J. Phys.: Condens. Matter}\ }\textbf {\bibinfo {volume}
  {27}},\ \bibinfo {pages} {243201} (\bibinfo {year} {2015})}\BibitemShut
  {NoStop}%
\bibitem [{\citenamefont {Tanaka}\ \emph {et~al.}(2012)\citenamefont {Tanaka},
  \citenamefont {Sato},\ and\ \citenamefont {Nagaosa}}]{Tanaka12}%
  \BibitemOpen
  \bibfield  {author} {\bibinfo {author} {\bibfnamefont {Y.}~\bibnamefont
  {Tanaka}}, \bibinfo {author} {\bibfnamefont {M.}~\bibnamefont {Sato}},\ and\
  \bibinfo {author} {\bibfnamefont {N.}~\bibnamefont {Nagaosa}},\ }\bibfield
  {title} {\bibinfo {title} {Symmetry and topology in superconductors
  odd-frequency pairing and edge states},\ }\href
  {https://doi.org/10.1143/JPSJ.81.011013} {\bibfield  {journal} {\bibinfo
  {journal} {J. Phys. Soc. Jpn.}\ }\textbf {\bibinfo {volume} {81}},\ \bibinfo
  {pages} {011013} (\bibinfo {year} {2012})}\BibitemShut {NoStop}%
\bibitem [{\citenamefont {Klitzing}\ \emph {et~al.}(1980)\citenamefont
  {Klitzing}, \citenamefont {Dorda},\ and\ \citenamefont
  {Pepper}}]{Klitzing80}%
  \BibitemOpen
  \bibfield  {author} {\bibinfo {author} {\bibfnamefont {K.~v.}\ \bibnamefont
  {Klitzing}}, \bibinfo {author} {\bibfnamefont {G.}~\bibnamefont {Dorda}},\
  and\ \bibinfo {author} {\bibfnamefont {M.}~\bibnamefont {Pepper}},\
  }\bibfield  {title} {\bibinfo {title} {{New Method for High-Accuracy
  Determination of the Fine-Structure Constant Based on Quantized Hall
  Resistance}},\ }\href {https://doi.org/10.1103/PhysRevLett.45.494} {\bibfield
   {journal} {\bibinfo  {journal} {Phys. Rev. Lett.}\ }\textbf {\bibinfo
  {volume} {45}},\ \bibinfo {pages} {494} (\bibinfo {year} {1980})}\BibitemShut
  {NoStop}%
\bibitem [{\citenamefont {Thouless}\ \emph {et~al.}(1982)\citenamefont
  {Thouless}, \citenamefont {Kohmoto}, \citenamefont {Nightingale},\ and\
  \citenamefont {den Nijs}}]{Thouless82}%
  \BibitemOpen
  \bibfield  {author} {\bibinfo {author} {\bibfnamefont {D.~J.}\ \bibnamefont
  {Thouless}}, \bibinfo {author} {\bibfnamefont {M.}~\bibnamefont {Kohmoto}},
  \bibinfo {author} {\bibfnamefont {M.~P.}\ \bibnamefont {Nightingale}},\ and\
  \bibinfo {author} {\bibfnamefont {M.}~\bibnamefont {den Nijs}},\ }\bibfield
  {title} {\bibinfo {title} {{Quantized Hall Conductance in a Two-Dimensional
  Periodic Potential}},\ }\href {https://doi.org/10.1103/PhysRevLett.49.405}
  {\bibfield  {journal} {\bibinfo  {journal} {Phys. Rev. Lett.}\ }\textbf
  {\bibinfo {volume} {49}},\ \bibinfo {pages} {405} (\bibinfo {year}
  {1982})}\BibitemShut {NoStop}%
\bibitem [{Note1()}]{Note1}%
  \BibitemOpen
  \bibinfo {note} {Mathematically, the Euler characteristic of a
  $d$-dimensional Fermi sea is defined as the alternating sum of its Betti
  numbers $b_k$, $\chi _F=\DOTSB \sum@ \slimits@ _{k=0}^d(-1)^k b_k$. See also
  \cite {TopoPhys}}\BibitemShut {NoStop}%
\bibitem [{\citenamefont {Lifshitz}(1960)}]{Lifshitz60}%
  \BibitemOpen
  \bibfield  {author} {\bibinfo {author} {\bibfnamefont {I.~M.}\ \bibnamefont
  {Lifshitz}},\ }\bibfield  {title} {\bibinfo {title} {{Anomalies of Electron
  Characteristics of a Metal in the High Pressure Region}},\ }\href
  {http://www.jetp.ras.ru/cgi-bin/e/index/e/11/5/p1130?a=list} {\bibfield
  {journal} {\bibinfo  {journal} {J. Exp. Theor. Phys.}\ }\textbf {\bibinfo
  {volume} {11}},\ \bibinfo {pages} {1130} (\bibinfo {year}
  {1960})}\BibitemShut {NoStop}%
\bibitem [{\citenamefont {Volovik}(2017)}]{Volovik17}%
  \BibitemOpen
  \bibfield  {author} {\bibinfo {author} {\bibfnamefont {G.~E.}\ \bibnamefont
  {Volovik}},\ }\bibfield  {title} {\bibinfo {title} {{Topological Lifshitz
  transitions}},\ }\href {https://doi.org/10.1063/1.4974185} {\bibfield
  {journal} {\bibinfo  {journal} {Low Temp. Phys.}\ }\textbf {\bibinfo {volume}
  {43}},\ \bibinfo {pages} {47} (\bibinfo {year} {2017})}\BibitemShut {NoStop}%
\bibitem [{\citenamefont {Kane}(2022)}]{Kane22}%
  \BibitemOpen
  \bibfield  {author} {\bibinfo {author} {\bibfnamefont {C.~L.}\ \bibnamefont
  {Kane}},\ }\bibfield  {title} {\bibinfo {title} {{Quantized Nonlinear
  Conductance in Ballistic Metals}},\ }\href
  {https://doi.org/10.1103/PhysRevLett.128.076801} {\bibfield  {journal}
  {\bibinfo  {journal} {Phys. Rev. Lett.}\ }\textbf {\bibinfo {volume} {128}},\
  \bibinfo {pages} {076801} (\bibinfo {year} {2022})}\BibitemShut {NoStop}%
\bibitem [{\citenamefont {Landauer}(1957)}]{Landauer57}%
  \BibitemOpen
  \bibfield  {author} {\bibinfo {author} {\bibfnamefont {R.}~\bibnamefont
  {Landauer}},\ }\bibfield  {title} {\bibinfo {title} {Spatial variation of
  currents and fields due to localized scatterers in metallic conduction},\
  }\href {https://doi.org/10.1147/rd.13.0223} {\bibfield  {journal} {\bibinfo
  {journal} {IBM J. Res. Dev.}\ }\textbf {\bibinfo {volume} {1}},\ \bibinfo
  {pages} {223} (\bibinfo {year} {1957})}\BibitemShut {NoStop}%
\bibitem [{\citenamefont {Tam}\ \emph {et~al.}(2022)\citenamefont {Tam},
  \citenamefont {Claassen},\ and\ \citenamefont {Kane}}]{Tam22}%
  \BibitemOpen
  \bibfield  {author} {\bibinfo {author} {\bibfnamefont {P.~M.}\ \bibnamefont
  {Tam}}, \bibinfo {author} {\bibfnamefont {M.}~\bibnamefont {Claassen}},\ and\
  \bibinfo {author} {\bibfnamefont {C.~L.}\ \bibnamefont {Kane}},\ }\bibfield
  {title} {\bibinfo {title} {Topological multipartite entanglement in a {F}ermi
  liquid},\ }\href {https://doi.org/10.1103/PhysRevX.12.031022} {\bibfield
  {journal} {\bibinfo  {journal} {Phys. Rev. X}\ }\textbf {\bibinfo {volume}
  {12}},\ \bibinfo {pages} {031022} (\bibinfo {year} {2022})}\BibitemShut
  {NoStop}%
\bibitem [{\citenamefont {Yang}\ and\ \citenamefont {Zhai}(2022)}]{Yang22QNL}%
  \BibitemOpen
  \bibfield  {author} {\bibinfo {author} {\bibfnamefont {F.}~\bibnamefont
  {Yang}}\ and\ \bibinfo {author} {\bibfnamefont {H.}~\bibnamefont {Zhai}},\
  }\bibfield  {title} {\bibinfo {title} {Quantized {N}onlinear {T}ransport with
  {U}ltracold {A}toms},\ }\href {https://doi.org/10.22331/q-2022-11-10-857}
  {\bibfield  {journal} {\bibinfo  {journal} {{Quantum}}\ }\textbf {\bibinfo
  {volume} {6}},\ \bibinfo {pages} {857} (\bibinfo {year} {2022})}\BibitemShut
  {NoStop}%
\bibitem [{\citenamefont {Zhang}(2023)}]{PFZhang23}%
  \BibitemOpen
  \bibfield  {author} {\bibinfo {author} {\bibfnamefont {P.}~\bibnamefont
  {Zhang}},\ }\bibfield  {title} {\bibinfo {title} {Quantized topological
  response in trapped quantum gases},\ }\href
  {https://doi.org/10.1103/PhysRevA.107.L031305} {\bibfield  {journal}
  {\bibinfo  {journal} {Phys. Rev. A}\ }\textbf {\bibinfo {volume} {107}},\
  \bibinfo {pages} {L031305} (\bibinfo {year} {2023})}\BibitemShut {NoStop}%
\bibitem [{\citenamefont {Tam}\ and\ \citenamefont {Kane}(2023)}]{Tam23}%
  \BibitemOpen
  \bibfield  {author} {\bibinfo {author} {\bibfnamefont {P.~M.}\ \bibnamefont
  {Tam}}\ and\ \bibinfo {author} {\bibfnamefont {C.~L.}\ \bibnamefont {Kane}},\
  }\bibfield  {title} {\bibinfo {title} {{Probing Fermi Sea Topology by Andreev
  State Transport}},\ }\href {https://doi.org/10.1103/PhysRevLett.130.096301}
  {\bibfield  {journal} {\bibinfo  {journal} {Phys. Rev. Lett.}\ }\textbf
  {\bibinfo {volume} {130}},\ \bibinfo {pages} {096301} (\bibinfo {year}
  {2023})}\BibitemShut {NoStop}%
\bibitem [{\citenamefont {Tam}\ \emph {et~al.}(2023)\citenamefont {Tam},
  \citenamefont {De~Beule},\ and\ \citenamefont {Kane}}]{tam2023topological}%
  \BibitemOpen
  \bibfield  {author} {\bibinfo {author} {\bibfnamefont {P.~M.}\ \bibnamefont
  {Tam}}, \bibinfo {author} {\bibfnamefont {C.}~\bibnamefont {De~Beule}},\ and\
  \bibinfo {author} {\bibfnamefont {C.~L.}\ \bibnamefont {Kane}},\ }\bibfield
  {title} {\bibinfo {title} {Topological {A}ndreev rectification},\ }\href
  {https://doi.org/10.1103/PhysRevB.107.245422} {\bibfield  {journal} {\bibinfo
   {journal} {Phys. Rev. B}\ }\textbf {\bibinfo {volume} {107}},\ \bibinfo
  {pages} {245422} (\bibinfo {year} {2023})}\BibitemShut {NoStop}%
\bibitem [{\citenamefont {Kitaev}(2006)}]{Kitaev06QSL}%
  \BibitemOpen
  \bibfield  {author} {\bibinfo {author} {\bibfnamefont {A.}~\bibnamefont
  {Kitaev}},\ }\bibfield  {title} {\bibinfo {title} {Anyons in an exactly
  solved model and beyond},\ }\href {https://doi.org/10.1016/j.aop.2005.10.005}
  {\bibfield  {journal} {\bibinfo  {journal} {Ann. Phys.}\ }\textbf {\bibinfo
  {volume} {321}},\ \bibinfo {pages} {2} (\bibinfo {year} {2006})}\BibitemShut
  {NoStop}%
\bibitem [{\citenamefont {Kane}\ and\ \citenamefont {Fisher}(1997)}]{Kane97}%
  \BibitemOpen
  \bibfield  {author} {\bibinfo {author} {\bibfnamefont {C.~L.}\ \bibnamefont
  {Kane}}\ and\ \bibinfo {author} {\bibfnamefont {M.~P.~A.}\ \bibnamefont
  {Fisher}},\ }\bibfield  {title} {\bibinfo {title} {Quantized thermal
  transport in the fractional quantum {H}all effect},\ }\href
  {https://doi.org/10.1103/PhysRevB.55.15832} {\bibfield  {journal} {\bibinfo
  {journal} {Phys. Rev. B}\ }\textbf {\bibinfo {volume} {55}},\ \bibinfo
  {pages} {15832} (\bibinfo {year} {1997})}\BibitemShut {NoStop}%
\bibitem [{\citenamefont {Cappelli}\ \emph {et~al.}(2002)\citenamefont
  {Cappelli}, \citenamefont {Huerta},\ and\ \citenamefont
  {Zemba}}]{Cappelli02}%
  \BibitemOpen
  \bibfield  {author} {\bibinfo {author} {\bibfnamefont {A.}~\bibnamefont
  {Cappelli}}, \bibinfo {author} {\bibfnamefont {M.}~\bibnamefont {Huerta}},\
  and\ \bibinfo {author} {\bibfnamefont {G.~R.}\ \bibnamefont {Zemba}},\
  }\bibfield  {title} {\bibinfo {title} {Thermal transport in chiral conformal
  theories and hierarchical quantum {H}all states},\ }\href
  {https://doi.org/10.1016/S0550-3213(02)00340-1} {\bibfield  {journal}
  {\bibinfo  {journal} {Nucl. Phys. B}\ }\textbf {\bibinfo {volume} {636}},\
  \bibinfo {pages} {568} (\bibinfo {year} {2002})}\BibitemShut {NoStop}%
\bibitem [{\citenamefont {Sato}(2009)}]{Sato09}%
  \BibitemOpen
  \bibfield  {author} {\bibinfo {author} {\bibfnamefont {M.}~\bibnamefont
  {Sato}},\ }\bibfield  {title} {\bibinfo {title} {Topological properties of
  spin-triplet superconductors and {F}ermi surface topology in the normal
  state},\ }\href {https://doi.org/10.1103/PhysRevB.79.214526} {\bibfield
  {journal} {\bibinfo  {journal} {Phys. Rev. B}\ }\textbf {\bibinfo {volume}
  {79}},\ \bibinfo {pages} {214526} (\bibinfo {year} {2009})}\BibitemShut
  {NoStop}%
\bibitem [{\citenamefont {Sato}(2010)}]{Sato10}%
  \BibitemOpen
  \bibfield  {author} {\bibinfo {author} {\bibfnamefont {M.}~\bibnamefont
  {Sato}},\ }\bibfield  {title} {\bibinfo {title} {Topological odd-parity
  superconductors},\ }\href {https://doi.org/10.1103/PhysRevB.81.220504}
  {\bibfield  {journal} {\bibinfo  {journal} {Phys. Rev. B}\ }\textbf {\bibinfo
  {volume} {81}},\ \bibinfo {pages} {220504(R)} (\bibinfo {year}
  {2010})}\BibitemShut {NoStop}%
\bibitem [{\citenamefont {Milnor}(1997)}]{TopoDiff}%
  \BibitemOpen
  \bibfield  {author} {\bibinfo {author} {\bibfnamefont {J.}~\bibnamefont
  {Milnor}},\ }\href
  {https://press.princeton.edu/books/paperback/9780691048338/topology-from-the-differentiable-viewpoint}
  {\emph {\bibinfo {title} {Topology from the Differentiable Viewpoint}}}\
  (\bibinfo  {publisher} {Princeton University Press},\ \bibinfo {year}
  {1997})\BibitemShut {NoStop}%
\bibitem [{\citenamefont {Read}\ and\ \citenamefont {Green}(2000)}]{Read00}%
  \BibitemOpen
  \bibfield  {author} {\bibinfo {author} {\bibfnamefont {N.}~\bibnamefont
  {Read}}\ and\ \bibinfo {author} {\bibfnamefont {D.}~\bibnamefont {Green}},\
  }\bibfield  {title} {\bibinfo {title} {Paired states of fermions in two
  dimensions with breaking of parity and time-reversal symmetries and the
  fractional quantum {H}all effect},\ }\href
  {https://doi.org/10.1103/PhysRevB.61.10267} {\bibfield  {journal} {\bibinfo
  {journal} {Phys. Rev. B}\ }\textbf {\bibinfo {volume} {61}},\ \bibinfo
  {pages} {10267} (\bibinfo {year} {2000})}\BibitemShut {NoStop}%
\bibitem [{\citenamefont {Kallin}\ and\ \citenamefont
  {Berlinsky}(2016)}]{Kallin16}%
  \BibitemOpen
  \bibfield  {author} {\bibinfo {author} {\bibfnamefont {C.}~\bibnamefont
  {Kallin}}\ and\ \bibinfo {author} {\bibfnamefont {J.}~\bibnamefont
  {Berlinsky}},\ }\bibfield  {title} {\bibinfo {title} {Chiral
  superconductors},\ }\href {https://doi.org/10.1088/0034-4885/79/5/054502}
  {\bibfield  {journal} {\bibinfo  {journal} {Rep. Prog. Phys.}\ }\textbf
  {\bibinfo {volume} {79}},\ \bibinfo {pages} {054502} (\bibinfo {year}
  {2016})}\BibitemShut {NoStop}%
\bibitem [{\citenamefont {Schnyder}\ \emph {et~al.}(2008)\citenamefont
  {Schnyder}, \citenamefont {Ryu}, \citenamefont {Furusaki},\ and\
  \citenamefont {Ludwig}}]{Schnyder08}%
  \BibitemOpen
  \bibfield  {author} {\bibinfo {author} {\bibfnamefont {A.~P.}\ \bibnamefont
  {Schnyder}}, \bibinfo {author} {\bibfnamefont {S.}~\bibnamefont {Ryu}},
  \bibinfo {author} {\bibfnamefont {A.}~\bibnamefont {Furusaki}},\ and\
  \bibinfo {author} {\bibfnamefont {A.~W.~W.}\ \bibnamefont {Ludwig}},\
  }\bibfield  {title} {\bibinfo {title} {Classification of topological
  insulators and superconductors in three spatial dimensions},\ }\href
  {https://doi.org/10.1103/PhysRevB.78.195125} {\bibfield  {journal} {\bibinfo
  {journal} {Phys. Rev. B}\ }\textbf {\bibinfo {volume} {78}},\ \bibinfo
  {pages} {195125} (\bibinfo {year} {2008})}\BibitemShut {NoStop}%
\bibitem [{\citenamefont {Kitaev}(2009)}]{Kitaev09}%
  \BibitemOpen
  \bibfield  {author} {\bibinfo {author} {\bibfnamefont {A.}~\bibnamefont
  {Kitaev}},\ }\bibfield  {title} {\bibinfo {title} {{Periodic table for
  topological insulators and superconductors}},\ }\href
  {https://doi.org/10.1063/1.3149495} {\bibfield  {journal} {\bibinfo
  {journal} {AIP Conf. Proc.}\ }\textbf {\bibinfo {volume} {1134}},\ \bibinfo
  {pages} {22} (\bibinfo {year} {2009})}\BibitemShut {NoStop}%
\bibitem [{\citenamefont {Chiu}\ \emph {et~al.}(2016)\citenamefont {Chiu},
  \citenamefont {Teo}, \citenamefont {Schnyder},\ and\ \citenamefont
  {Ryu}}]{Chiu16}%
  \BibitemOpen
  \bibfield  {author} {\bibinfo {author} {\bibfnamefont {C.-K.}\ \bibnamefont
  {Chiu}}, \bibinfo {author} {\bibfnamefont {J.~C.~Y.}\ \bibnamefont {Teo}},
  \bibinfo {author} {\bibfnamefont {A.~P.}\ \bibnamefont {Schnyder}},\ and\
  \bibinfo {author} {\bibfnamefont {S.}~\bibnamefont {Ryu}},\ }\bibfield
  {title} {\bibinfo {title} {Classification of topological quantum matter with
  symmetries},\ }\href {https://doi.org/10.1103/RevModPhys.88.035005}
  {\bibfield  {journal} {\bibinfo  {journal} {Rev. Mod. Phys.}\ }\textbf
  {\bibinfo {volume} {88}},\ \bibinfo {pages} {035005} (\bibinfo {year}
  {2016})}\BibitemShut {NoStop}%
\bibitem [{\citenamefont {Schwarz}(1994)}]{TopoPhys}%
  \BibitemOpen
  \bibfield  {author} {\bibinfo {author} {\bibfnamefont {A.~S.}\ \bibnamefont
  {Schwarz}},\ }\href {https://doi.org/10.1007/978-3-662-02998-5} {\emph
  {\bibinfo {title} {{Topology for Physicists}}}}\ (\bibinfo  {publisher}
  {Springer-Verlag Berlin Heidelberg},\ \bibinfo {year} {1994})\BibitemShut
  {NoStop}%
\bibitem [{Note2()}]{Note2}%
  \BibitemOpen
  \bibinfo {note} {In the following, we assume all zeros are isolated. When
  there are lines of zeros, one can always smoothly change the vector field by
  an infinitesimal amount such that all zeros are isolated.}\BibitemShut
  {Stop}%
\bibitem [{Note3()}]{Note3}%
  \BibitemOpen
  \bibinfo {note} {Eq. (\ref {eq:index}) is equivalent to the Morse theory
  formulation of $\chi _F$. $\nu _i({\protect \bf v})$ equals to the sign of
  the Hessian of $\epsilon _{\protect \bf k}$ at each critical
  point.}\BibitemShut {Stop}%
\bibitem [{\citenamefont {Hatcher}(2002)}]{Hatcher}%
  \BibitemOpen
  \bibfield  {author} {\bibinfo {author} {\bibfnamefont {A.}~\bibnamefont
  {Hatcher}},\ }\href
  {https://www.cambridge.org/gb/academic/subjects/mathematics/geometry-and-topology/algebraic-topology-1?format=PB&isbn=9780521795401}
  {\emph {\bibinfo {title} {Algebraic Topology}}}\ (\bibinfo  {publisher}
  {Cambridge University Press},\ \bibinfo {year} {2002})\BibitemShut {NoStop}%
\bibitem [{\citenamefont {Volovik}(1988)}]{Volovik88}%
  \BibitemOpen
  \bibfield  {author} {\bibinfo {author} {\bibfnamefont {G.}~\bibnamefont
  {Volovik}},\ }\bibfield  {title} {\bibinfo {title} {{Quantized Hall effect in
  superfluid Helium-3 film}},\ }\href
  {https://doi.org/10.1016/0375-9601(88)90373-8} {\bibfield  {journal}
  {\bibinfo  {journal} {Phys. Lett. A}\ }\textbf {\bibinfo {volume} {128}},\
  \bibinfo {pages} {277} (\bibinfo {year} {1988})}\BibitemShut {NoStop}%
\bibitem [{\citenamefont {Leggett}(1975)}]{Leggett75}%
  \BibitemOpen
  \bibfield  {author} {\bibinfo {author} {\bibfnamefont {A.~J.}\ \bibnamefont
  {Leggett}},\ }\bibfield  {title} {\bibinfo {title} {A theoretical description
  of the new phases of liquid $^{3}\mathrm{He}$},\ }\href
  {https://doi.org/10.1103/RevModPhys.47.331} {\bibfield  {journal} {\bibinfo
  {journal} {Rev. Mod. Phys.}\ }\textbf {\bibinfo {volume} {47}},\ \bibinfo
  {pages} {331} (\bibinfo {year} {1975})}\BibitemShut {NoStop}%
\bibitem [{\citenamefont {Volovik}(2009)}]{Volovik}%
  \BibitemOpen
  \bibfield  {author} {\bibinfo {author} {\bibfnamefont {G.~E.}\ \bibnamefont
  {Volovik}},\ }\href
  {https://doi.org/10.1093/acprof:oso/9780199564842.001.0001} {\emph {\bibinfo
  {title} {{{The Universe in a Helium Droplet}}}}}\ (\bibinfo  {publisher}
  {Oxford University Press},\ \bibinfo {year} {2009})\BibitemShut {NoStop}%
\bibitem [{\citenamefont {Vollhardt}\ and\ \citenamefont
  {W{\"o}lfle}(1990)}]{Vollhardt}%
  \BibitemOpen
  \bibfield  {author} {\bibinfo {author} {\bibfnamefont {D.}~\bibnamefont
  {Vollhardt}}\ and\ \bibinfo {author} {\bibfnamefont {P.}~\bibnamefont
  {W{\"o}lfle}},\ }\href {https://doi.org/10.1201/b12808} {\emph {\bibinfo
  {title} {{The Superfluid Phases of Helium 3}}}}\ (\bibinfo  {publisher}
  {Taylor \& Francis},\ \bibinfo {year} {1990})\BibitemShut {NoStop}%
\bibitem [{\citenamefont {Fu}\ and\ \citenamefont {Kane}(2008)}]{Fu08}%
  \BibitemOpen
  \bibfield  {author} {\bibinfo {author} {\bibfnamefont {L.}~\bibnamefont
  {Fu}}\ and\ \bibinfo {author} {\bibfnamefont {C.~L.}\ \bibnamefont {Kane}},\
  }\bibfield  {title} {\bibinfo {title} {{Superconducting Proximity Effect and
  Majorana Fermions at the Surface of a Topological Insulator}},\ }\href
  {https://doi.org/10.1103/PhysRevLett.100.096407} {\bibfield  {journal}
  {\bibinfo  {journal} {Phys. Rev. Lett.}\ }\textbf {\bibinfo {volume} {100}},\
  \bibinfo {pages} {096407} (\bibinfo {year} {2008})}\BibitemShut {NoStop}%
\bibitem [{\citenamefont {Abrikosov}(1988)}]{Abrikosov}%
  \BibitemOpen
  \bibfield  {author} {\bibinfo {author} {\bibfnamefont {A.~A.}\ \bibnamefont
  {Abrikosov}},\ }\href@noop {} {\emph {\bibinfo {title} {Fundamentals of the
  Theory of Metals}}}\ (\bibinfo  {publisher} {North-Holland},\ \bibinfo
  {address} {Amsterdam},\ \bibinfo {year} {1988})\BibitemShut {NoStop}%
\bibitem [{\citenamefont {Yang}\ \emph {et~al.}(2019)\citenamefont {Yang},
  \citenamefont {Jiang},\ and\ \citenamefont {Zhou}}]{Yang19}%
  \BibitemOpen
  \bibfield  {author} {\bibinfo {author} {\bibfnamefont {F.}~\bibnamefont
  {Yang}}, \bibinfo {author} {\bibfnamefont {S.-J.}\ \bibnamefont {Jiang}},\
  and\ \bibinfo {author} {\bibfnamefont {F.}~\bibnamefont {Zhou}},\ }\bibfield
  {title} {\bibinfo {title} {Robust cusps near topological phase transitions:
  Signatures of {M}ajorana fermions and interactions with fluctuations},\
  }\href {https://doi.org/10.1103/PhysRevB.100.054508} {\bibfield  {journal}
  {\bibinfo  {journal} {Phys. Rev. B}\ }\textbf {\bibinfo {volume} {100}},\
  \bibinfo {pages} {054508} (\bibinfo {year} {2019})}\BibitemShut {NoStop}%
\bibitem [{\citenamefont {Yang}\ and\ \citenamefont {Zhou}(2021)}]{Yang21}%
  \BibitemOpen
  \bibfield  {author} {\bibinfo {author} {\bibfnamefont {F.}~\bibnamefont
  {Yang}}\ and\ \bibinfo {author} {\bibfnamefont {F.}~\bibnamefont {Zhou}},\
  }\bibfield  {title} {\bibinfo {title} {Topological quantum criticality in
  superfluids and superconductors: Surface criticality, thermal properties, and
  {L}ifshitz {M}ajorana fields},\ }\href
  {https://doi.org/10.1103/PhysRevB.103.205126} {\bibfield  {journal} {\bibinfo
   {journal} {Phys. Rev. B}\ }\textbf {\bibinfo {volume} {103}},\ \bibinfo
  {pages} {205126} (\bibinfo {year} {2021})}\BibitemShut {NoStop}%
\bibitem [{\citenamefont {Zhou}(2022)}]{Zhou22}%
  \BibitemOpen
  \bibfield  {author} {\bibinfo {author} {\bibfnamefont {F.}~\bibnamefont
  {Zhou}},\ }\bibfield  {title} {\bibinfo {title} {Topological quantum critical
  points in strong coupling limits: Global symmetries and strongly interacting
  {M}ajorana fermions},\ }\href {https://doi.org/10.1103/PhysRevB.105.014503}
  {\bibfield  {journal} {\bibinfo  {journal} {Phys. Rev. B}\ }\textbf {\bibinfo
  {volume} {105}},\ \bibinfo {pages} {014503} (\bibinfo {year}
  {2022})}\BibitemShut {NoStop}%
\bibitem [{\citenamefont {Yerin}\ \emph {et~al.}(2022)\citenamefont {Yerin},
  \citenamefont {Varlamov},\ and\ \citenamefont {Petrillo}}]{Yerin22}%
  \BibitemOpen
  \bibfield  {author} {\bibinfo {author} {\bibfnamefont {Y.}~\bibnamefont
  {Yerin}}, \bibinfo {author} {\bibfnamefont {A.~A.}\ \bibnamefont
  {Varlamov}},\ and\ \bibinfo {author} {\bibfnamefont {C.}~\bibnamefont
  {Petrillo}},\ }\bibfield  {title} {\bibinfo {title} {Topological nature of
  the transition between the gap and the gapless superconducting states},\
  }\href {https://doi.org/10.1209/0295-5075/ac64b9} {\bibfield  {journal}
  {\bibinfo  {journal} {Europhys. Lett.}\ }\textbf {\bibinfo {volume} {138}},\
  \bibinfo {pages} {16005} (\bibinfo {year} {2022})}\BibitemShut {NoStop}%
\bibitem [{\citenamefont {Qi}\ \emph {et~al.}(2010)\citenamefont {Qi},
  \citenamefont {Hughes},\ and\ \citenamefont {Zhang}}]{Qi10}%
  \BibitemOpen
  \bibfield  {author} {\bibinfo {author} {\bibfnamefont {X.-L.}\ \bibnamefont
  {Qi}}, \bibinfo {author} {\bibfnamefont {T.~L.}\ \bibnamefont {Hughes}},\
  and\ \bibinfo {author} {\bibfnamefont {S.-C.}\ \bibnamefont {Zhang}},\
  }\bibfield  {title} {\bibinfo {title} {Topological invariants for the {F}ermi
  surface of a time-reversal-invariant superconductor},\ }\href
  {https://doi.org/10.1103/PhysRevB.81.134508} {\bibfield  {journal} {\bibinfo
  {journal} {Phys. Rev. B}\ }\textbf {\bibinfo {volume} {81}},\ \bibinfo
  {pages} {134508} (\bibinfo {year} {2010})}\BibitemShut {NoStop}%
\bibitem [{\citenamefont {Zhang}\ \emph {et~al.}(2020)\citenamefont {Zhang},
  \citenamefont {Zhang},\ and\ \citenamefont {Liu}}]{Zhang20}%
  \BibitemOpen
  \bibfield  {author} {\bibinfo {author} {\bibfnamefont {L.}~\bibnamefont
  {Zhang}}, \bibinfo {author} {\bibfnamefont {L.}~\bibnamefont {Zhang}},\ and\
  \bibinfo {author} {\bibfnamefont {X.-J.}\ \bibnamefont {Liu}},\ }\bibfield
  {title} {\bibinfo {title} {{Unified Theory to Characterize Floquet
  Topological Phases by Quench Dynamics}},\ }\href
  {https://doi.org/10.1103/PhysRevLett.125.183001} {\bibfield  {journal}
  {\bibinfo  {journal} {Phys. Rev. Lett.}\ }\textbf {\bibinfo {volume} {125}},\
  \bibinfo {pages} {183001} (\bibinfo {year} {2020})}\BibitemShut {NoStop}%
\end{thebibliography}%

\end{document}